\documentclass[letterpaper,12pt]{article}
\usepackage{graphicx}  
\usepackage{cite}

\def\ab{$ab$ $initio$}

\begin{document}

\bibliographystyle{AIP}

\renewcommand{\thefootnote}{\fnsymbol{footnote}}

\begin{center}
{\bf Sealing off a carbon nanotube with a self-assembled aqueous valve \\
for the storage of hydrogen in GPa pressure}\\

\bigskip
\bigskip
H.Y.~Chen$^{1,3}$, D.Y. Sun$^{1}$, X.G.~Gong,$^2$, and
Zhi-Feng~Liu$^3$\footnote{corresponding authors. email:
zfliu@cuhk.edu.hk}           \\
\medskip
$^1$Department of Physics \\
East China Normal University, Shanghai 200062, China \\
\bigskip
and \\
$^2$Surface Physics Laboratory and Department of Physics \\
Fudan University, Shanghai 200433, China \\
and \\
$^3$Department of Chemistry and \\
Centre for Scientific Modeling and Computation \\
Chinese University of Hong Kong \\
Shatin, Hong Kong, China \\
\bigskip

\end{center}

\noindent{\bf Abstract:}

The end section of a carbon nanotube, cut by acid treatment,
contains hydrophillic oxygen groups.  Water molecules can
self-assemble around these groups to seal off a carbon nanotube
and form an ``aqueous valve''.  Molecular dynamics simulations
on single-wall (12,12) and (15,15) tubes with dangling
carboxyl groups show that the formation of aqueous valves
can be achieved both in the absence of and in the presence of
high pressure hydrogen.  Furthermore,
significant diffusion barriers through aqueous valves
are identified.  It indicates that such valves could
hold hydrogen inside the tube with GPa pressure.
Releasing hydrogen is easily achieved by melting the
``aqueous valve''.  Such a design provides a recyclable and
non-destructive way to store hydrogen in GPa pressure.
Under the storage conditions dictated by sealing off
the container in liquid water, the hydrogen density inside
the container is higher than that for solid hydrogen,
which promises excellent weight storage efficiency.

\vfill\eject

As cylinders formed by rolling up robust graphenes,
carbon nanotubes enclose much empty spaces.   Foreign particles,
including both ions and molecules, can be inserted into these tubes,
to form quasi-1D structures with interesting physical and chemical
properties.\cite{Monthioux02,GreenMH:acr02,KhlobystovBB05} Inserting
hydrogen molecules into carbon nanotubes can be easily achieved, due
to their small size.  But for hydrogen, it is the compressibility,
rather than the structure, that is the most
interesting,\cite{YeGGSL07,PupyshevaFY08,SunCLGL09,ChenGLS11,ChenLGS11:2,SuyetinV11}
since H$_2$ has a weak van der Waals potential and is
highly compressible all the way up to
and above GPa pressure (tens of thousands of
bars).\cite{Silvera:h2solid,Mao:h2order} The mechanical strength of
a carbon nanotube could stand pressure up to 40 GPa, as measured by
experiment.\cite{SunBKRTA06} Therefore, compressing hydrogen to
GPa pressure, which is two orders of magnitude higher than the pressure
in a typical gas tank, would be ideal for the storage of
hydrogen, provided a suitable nano-scale valve could
be made.\cite{YeGGSL07,PupyshevaFY08,SunCLGL09,SuyetinV11}
Such a valve should block hydrogen inside carbon nanotubes,
and control the filling and release of hydrogen in a recyclable
and non-destructive way, which is a
challenge in both design and implementation.

Attempts have been made to design such a nano valve.
Our groups have suggested a design modeled
after the "ball check valve", composed of
a C$_{60}$ molecule enclosed by a half-spherical carbon nanotube
cap. Molecular dynamics (MD) simulations indicated that the position
of C$_{60}$ could be switched between ``open'' and ''close'' states,
as the external pressure was varied.\cite{SunCLGL09,YeGGSL07}
Vakhrushev and Suyetin suggested that a positively charged
endohedral complex K$^+$@C$_{60}$ could be used as the blocker for
the storage of hydrogen and methane, which could then be controlled
by the application of electric field.\cite{SuyetinV11,VakhrushevS09}
Transition metal particles are often attached to the end
of carbon nanotubes during the growth process, and asymmetric
permeability of hydrogen through metal layers could also
offer opportunities in designing a nano valve.\cite{Yakobson87}
However, turning such elaborate designs into actual structures at
the molecular level would be quite challenging.
In this paper, we present a new design
based on readily made structures that could be easily
implemented.

Our starting point is the common structural feature
of a carbon nanotube, typically consisted of two parts:
a cylindrical tube body, which nowadays could
be made into considerable length with little
defects, and two end sections, where the caps are usually cut off by
acid treatment and the dangling bonds are typically saturated by
oxygen containing groups, such as carboxyl(-COOH), carbonyl (-C=O)
or hydroxy groups(-OH).\cite{Yates:o-tube2,Yates:tubeh2so4}  As
curved graphene sheet, the tube wall contains no polar groups, and
would prefer to bind non-polar molecules such as hydrogen.  In
contrast, the oxygen containing groups at the end sections are
polar, and would prefer to bind with water molecules through
hydrogen bonds.

Our design for the valve is to seal the open end section by exposure
to water molecules.  Around oxygen containing hydrophilic groups,
water molecules should automatically aggregate
to form an ``aqueous valve'', hold together by hydrogen bonds.
There must be a barrier for the diffusion of hydrogen through an
aqueous valve.  Filling hydrogen into the tube would require the growth
of an aqueous valve in the presence of hydrogen under high
pressure, while the diffusion barrier should
lock the hydrogen inside after the withdrawal
of external pressure. Experimental implementation of such a valve is
straightforward, as it depends on the self-assembly of water
molecules around the hydrophilic groups.  Furthermore, nor is it
difficult to modify these groups so that their structures and
hydrophilic propensities could be adjusted. The crucial questions are
therefore to examine the growth of the aqueous
valve and to understand the mechanism and barrier for hydrogen
diffusion through such a valve at various pressures.

Although the energy of one hydrogen bond between two water molecules
is only around 5 kcal/mol, it has been demonstrated
by experiments that hydrogen molecules squeezed into ice
under GPa pressure remained locked inside ice lattice
when the external pressure was withdrawn.\cite{StruzhkinMMMH07,MaoMMEHCCSH06}
Even at GPa pressure, the hydrogen bonded ice lattice
could stand up to the the van der Waals repulsion
exerted by hydrogen molecules.  The relative weakness
of the hydrogen bonds, compared to typical
chemical bonds, is actually a big advantage when it comes to the
release of hydrogen through an aqueous valve.  Moderate heating
above the melting point easily disrupts the hydrogen bonds and opens
the valves.  Furthermore, such heating does not damage the dangling
groups, and aqueous valves can be formed again upon
exposure to water at lower temperatures.

For this study, a molecular model is built as shown
in Figure~1a.  Two types of singled walled carbon nanotubes,
with an index of (12,12) or (15,15) respectively, are studied.
Each tube is capped at one end and functionalized
by carboxyl groups (-COOH) on the other end.  The
structure shall be labeled as (12,12)-COOH (with 24 carboxyl
groups) and (15,15)-COOH (with 30
carboxyl groups).

Aggregation of water molecules around the carboxyl
groups is readily observed in the following
simulation.  Initially, a chunk of regular ice of Ih
lattice, containing 640 water molecules, is placed
8 \AA\ below the carboxyl groups, as shown in Figure~1b.
During an MD simulation of 10 ns at 220 K,
the ice chunk is melted and the water molecules
are self-assembled around the hydrophilic carboxyl
groups.  The temperature is then gradually lowered from
220 K to 77 K over 1 ns, and an
amorphous structure of water aggregates
is formed at the end section, as shown in Figure~1c, which blocks
the entrance into the tube and could potentially
be an aqueous valve.  Such a self-assembly
process is facilitated when water is in its liquid
state, and the simulation temperature of 220 K
is above the melting point of water as described
by the SPC/E potential, which is around 215 K.\cite{AlexiadisK08}

It is known from the phase diagram
of ice\cite{IcePhysics} that its melting point changes little with
pressure.  The triple point for ice V, ice VI, and liquid water
is identified at 0.63 GPa and 273 K.  It means that
in experiment hydrogen gas could be pressed to 0.63 GPa
just above 273 K to fill open nanotube containers, and
exposure to water, which is in liquid state under such
conditions, could then induce self-assembly
around the end section to seal off these containers.
Such a scheme could also work at higher pressure,
to $\sim$1 GPa, when the temperature is
raised to $\sim$ 300 K, according to the experimental
phase diagram.

For such a filling process, one must ask whether the
formation of an aqueous valve is affected by the presence
of hydrogen under high pressure.  To simulate it,
two artificial plates are introduced outside the tube,
as shown in Figure~2a.  Space between the two plates
are filled by hydrogen molecules.  By placing the bottom plate
at different positions, the pressure of hydrogen
could be raised to GPa range.  Again, a chunk of
ice is placed below the carboxyl groups, leaving
the end section open for the filling of hydrogen
inside the tube.

In the beginning step, water molecules in the ice chunk
are artificially frozen in their positions, as hydrogen molecules
are introduced between the plates at 250 K, which
diffuse not only into the tube but also into the lattice
of the fixed ice chunk, eventually reaching
an equilibrium pressure of 0.6 GPa in 300 ps.  Then the frozen water
molecules are set free, and the temperature is raised
to 270 K, so that the ice chunk is melted and the
self-assembly of water molecules around the hydrophilic carboxyl
groups is started.  After 300 ps, an aqueous valve is
again formed.  The system is then cooled down to
77 K in 500 ps, and the process produces the same
kind of aqueous valve as shown in Figure~1,
obtained in the absence of hydrogen.  Careful examination of the
structure shows that no hydrogen molecule is left inside
the water aggregate.  The result is not surprising because
the non-polar and hydrophobic hydrogen molecules do not mix
with the polar water molecules, and with their weak van der
Waals potentials, hydrogen molecules can easily move out
of the way of the aggregating water molecules.
Finally, the two plates and the hydrogen molecules
outside the sealed container are withdrawn,
as in Figure~2b, and the system is
equilibrated at 77 K for 300 ps.
The aqueous valve is stable and no leakage
of hydrogen molecules from inside the tube, which is at
0.6 GPa, is observed during the simulation.


While the above simulations prove that the container
could be sealed off, with and without the presence of
high pressure hydrogen,
the robustness of such an aqueous valve must be further
examined by calculating the diffusion barriers.
A number of calculations are carried out in which a hydrogen molecule
is forced to move through an aqueous valve by constraining
the $z$ coordinate of one particular hydrogen molecule, with
$z$ axis being the axial direction of the carbon nanotube.
Typically, this hydrogen molecule would move close to
the tube wall and find a path near the wall.  As there
is no hydrogen bond between the tube wall and
water molecules, such a diffusion path
produces less structural disruption.

Shown in Figure~3a is a typical potential energy
curve along the diffusion path, when the internal pressure inside
the tube and the external pressure outside the
tube are both at zero.  The $z$ coordinate is chosen so that
the averaged $z$ coordinate for the C=O oxygen atoms in
the -COOH groups around the end section is zero.  In
other words, $z=0$ could be treated as an approximate
dividing line, so that a hydrogen molecule is inside the container
at positive $z$ and outside the container at negative
$z$ values.  The energy at $z=$10 \AA\ (inside the tube)
is $\sim$ 0.1 eV lower than the energy at $z=$-20 \AA\ (outside
the tube), since a hydrogen molecule
inside the tube has more favorable van der Waals interactions
with carbon atoms on the tube wall and other hydrogen molecules
than its interaction with water molecules at an
interstitial site of ice.

There are many barriers and wells connecting these two points along
the potential energy curve.
Starting from $z=$10 \AA, when the hydrogen moves into
the aqueous valve, it passes through two interstitial
sites, seen as two energy minima (around $z=$1 and $z=4$ \AA\
respectively), and hits a barrier of 1.0 eV centered near $z=$0,
where the carboxyl groups are located.
Inside the tube, the diffusing hydrogen passes through the space
between water and the nanotube wall.  Near the end
of the tube, the hydrogen molecule must squeezes
through the spaces around the carboxyl groups covalently
bonded to the nanotube, which accounts for the big
barrier.  The energy difference between the top
of this peak and the position inside the tube is
defined as the release barrier.  Afterwards,
the hydrogen moves into the outer part of the
aqueous valve, which is mainly amorphous ice.  There
are several barriers in this region, which are
more or less comparable to the diffusion barrier in
Ih ice ($\sim$0.3 eV), although it could bump into
a larger barrier too (e.g. at -12 \AA), which is due
to the disorder in the ice structure.
There are of course many possible
trajectories, and shown in Figure~3a is only a typical path.
Molecular vibrations among the water molecules, carboxyl groups, and
carbon nanotubes would produce further fluctuation in such paths.
For each set of external and internal pressure,
we have optimized eight diffusion paths
and obtained average release barrier, listed
in Table~1.

For (12,12)-COOH, the average release barrier decreases
only slightly as the internal pressure is raised from 0
to 2 GPa.  These values around 0.8 eV are
significantly higher than the diffusion barrier
through Ih ice ($\sim$0.3 eV), indicating that
hydrogen can be held up inside the tube.
The slight decrease in the release barrier is
in agreement with the observation that the
increase in internal pressure only induces
small changes in the radial distribution function
(RDF) of water molecules, shown in Figure~3b and 3c.
When both $P_{ext}$ and $P_{int}$ are at zero,
the RDF around the -COOH groups at the end
section and the RDF in the amorphous ice outside the container are
almost identical, with a sharp peak around 2.64 \AA\ and a broad
feature centered just above 4 \AA.
When the internal pressure $P_{int}$ is increased to 1
GPa and the external pressure maintained at 0, the
broad feature centered around 4 \AA\ is maintained
with its center slightly shifted to shorter distance
for the RDF around the -COOH groups.  For the ice
outside the tube, the RDF is
little changed.  In other words, there is only slight
structural tightening around the -COOH groups and
little effect on the ice outside of the tube,
when $P_{int}$ is raised to 1 GPa, as the effect of
the internal pressure is restricted by the tube
wall and the valve.

The release barriers for the (15,15)-COOH valve are also
calculated, as listed in Table 1, and the values
are typically lower than
the corresponding values for the (12,12)-COOH valve.
More significant changes in the release barrier are
also observed when the pressure is varied.  With
$P_{ext}$ kept at zero, there is still a substantial
barrier above 0.6 eV for the release of hydrogen, when $P_{int}$ is
at 0 or 1 GPa.  However, when $P_{int}$ is raised further to 2.0
GPa, the barrier drops to around 0.3 eV,
which is comparable to the calculated diffusion barrier in ice.  It
indicates the difficulty for the (15,15)-COOH valve to hold hydrogen at an
internal pressure of 2 GPa.  Such differences between the
(15,15)-COOH and (12,12)-COOH valves are due to the difference
in their diameters, 20.3 \AA\ for the former, and 16.3 \AA\ for the latter,
respectively.  For the larger (15,15)-COOH valve, the end section
is not completely dominated by the -COOH groups at the edge and their hydrogen
bond interactions with the water molecules, because there is a
chunk of ice around the center of the section.  It makes the
valve more flexible as a hydrogen molecule squeeze through
the valve, which accounts for reduced barriers for hydrogen
diffusion.  One could project that as the
tube diameter increases further, the influence exerted by
the -COOH groups will be limited only to the edge of the
valve, and most part of the valve is just a chunk of ice.
To seal off such tubes for hydrogen storage, the length
of the hydrophilic groups should be extended, to tighten
their hold around the water aggregate.


The carboxyl groups, on which an aqueous valve is
assembled, contain C=O, C-O and O-H bonds, which
are robust covalent bonds little affected by the pressure of a few
GPa.  The van der Waals repulsions only slightly
decrease the average bond distances, by $\sim$0.001 \AA\
for the C-O bond and $\sim$ 0.0001 \AA\ for C=O bond,
when the internal and external pressures are raised
from 0 to 2 GPa.

The weak structural link in our model aqueous valve is
the hydrogen bonds that hold the water aggregate to the
carboxyl groups.  The energy for such a typical hydrogen
bond is only 5 kcal/mol, and at high internal pressure, the
force exerted on the valve may be strong enough to
break collectively these hydrogen bonds and push the
water aggregate out of the tube.  The threshold pressure
can be evaluated by molecular dynamics simulations.  Hydrogen
molecules are added consecutively inside a sealed container step by
step.  At each step, the configuration is first optimized
at 0 K, and then the temperature is gradually raised to
77 K and equilibrated for 100 ps.  The threshold pressure
is found to be 3.0 GPa for (12,12)-COOH and 2.6 GPa for
(15,15)-COOH respectively, when the aqueous valve is
pushed out during the equilibration step

The storage pressure envisioned for aqueous valves
is between 0.6 GPa around 273 K and 1.0 GPa around 300 K,
for which the mechanical stability is not a concern.
These two pressures are determined by the phase
diagram of water.  To seal off the valve, water should be in
liquid state to facilitate its self-assembly around the
tube ends.  Using the equation of state measured by Mills
and co-workers,\cite{MillsLBS77} the molar
volume of hydrogen is 17.1693 cm$^3$
at 0.6 GPa and 273 K, and 14.6072 cm$^3$ at 1.0 GPa and 300 K,
both being more favorable than even the molar
volume around 23 cm$^3$ for solid hydrogen at 4.2 K.\cite{Silvera:h2solid}

The weight storage efficiency varies with the tube diameter.
With the carbon atoms on the tube surface, the efficiency
is higher for tubes with larger diameters since they have larger
volume to surface ratios.  For a long tube, the weight of
the aqueous valve adds only a very small overhead and could
be ignored, and the H/C mass ratio $\alpha$
can be estimated by the weight of hydrogen contained in the
volume of a tube segment and the weight of carbon on the tube
wall (see supporting information), which is

$$\alpha = {3d_{CC}\pi \rho \over 48n} ({1.23n \over \pi}-d_{CH}  )^2 $$

\noindent for an (n,0) tube, and

$$\alpha = {\sqrt{3} d_{CC}\pi \rho \over 48n} ({1.23 \sqrt{3} n \over \pi}-d_{CH}  )^2 $$

\noindent for an (n,n) tube.  In the above formula, $d_{CC}$ is the
C--C distance on a carbon nanotube, taken to be 1.42 \AA, $\rho$ is the hydrogen
mass density in a.m.u/\AA$^3$, and $d_{CH}$ is the shortest distance
between the surface of a spherical hydrogen molecule and a carbon atom
on a carbon nanotube for the calculation of available storage volume,
which is taken as 1.6 \AA\ based on previous studies.\cite{WangJ99}
At 0.6 GPa and 273 K, the calculated weight efficiency, which is plotted
in Figure~4, reaches 5$\%$ at a tube diameter
of 2.0 nm and 8$\%$ at 3.0 nm.  At 1 GPa
and 300 K, the weight efficiency reaches 5$\%$ at 1.8 nm,
and close to 10$\%$ around 3.0 nm.

In principle, an aqueous valve
could be built for larger tubes by extending the -COOH
groups around the tube edge in our model to longer polymeric
and hydrophillic chains so that they could be extended to
the central potion of the valve.  The melting point of an aqueous
valve is another important consideration.  While the storage
of high pressure hydrogen inside the tube requires the aqueous
valve to be in the solid state, the release of the hydrogen is
achieved by the melting of the aqueous valve.  It should be noted
that many opportunities exist in modifying the functional groups to
produce various structures for the aqueous valve and to adjust its
melting point.  In synthetic chemistry, the oxygen containing groups can
be easily transformed.  The length, ionic strength and branching can be
controlled by introducing groups that bound more strongly with water
molecules, such as polycarbonate, polysilicate,
or polyphosphate groups.  The change in the pH value or the presence of salts
could also affect strength of the binding of these groups to
water.  These aspects of the valve design are currently being
investigated in our groups.

In summary, we have demonstrated by molecular dynamics simulation
that the oxygen containing dangling groups around the edge of an
opened carbon nanotube could serve as the frame for the self-assembly
of water molecules to produce an aqueous valve.  Formation of
such an aqueous valve is also demonstrated in
the presence of high pressure H$_2$.
Using a (12,12) tube and a (15,15) tube with -COOH groups as models,
substantial diffusion barriers are identified, for storage pressure
in the 1-2 GPa range.  They provide the basis for a new design of
high pressure nano-containers for the storage of hydrogen that could
be implemented and tested in experiments.

\bigskip
\noindent{\bf Computational Details:}

Molecular dynamics simulations are performed
by the TINKER 4.2 package.\cite{PonderR87}
The equation of
motion is solved by the Beeman algorithm with time
step of 1 fs\cite{Allen:MDbook} and the system temperature is controlled
by a Nos\'e-Hoover thermostat.\cite{Nose:stat,Hoover:stat}  The OPLS-all-atom field
with L-J potentials is applied to the carbon atoms on
CNT and to the carboxyl groups.  For water molecules,
SPC/E model is employed.  A hydrogen molecule is treated
as spherical and the interactions between two hydrogen
molecules are modeled by the Silvera-Goldman
potential.\cite{Silvera:h2potential,Silvera:h2solid}
The interaction between hydrogen and water molecules
is L-J potential.\cite{AlaviRK05}  Our calculated
barriers for type-sII clathrate are in good agreement
with the previously reported values.\cite{FrankcombeK07}
Within Ih ice, the diffusion barrier of a hydrogen
molecule is found to be 0.3 eV along $a/b$ direction
and 0.2 eV along $c$ direction, which is also
reproduced by simulations using the LAMMPS (Large-scale Atomic/Molecular
Massively Parallel Simulator) program.\cite{Plimpton95}
The interactions between carbon atoms
are modeled by the second-generation reactive empirical
bond order potential.\cite{BrennerSHSNS02}
The van der Waals interaction between a hydrogen molecule
and a carbon atom is described by the recently fitted
potential by some of us, which reproduces the high level
\ab\ results over a wide range of attractive and
repulsive regions.\cite{SunLGL07}  The cutoff for all unbonded
interactions is set to 60 \AA.  For different types
of atoms, the L-J potential is calculated by the
Lorentz-Berthlot combination rules.  Four carbon
atoms on the carbon nanotube are fixed
to prevent unnecessary movement of the tube
during all the simulations.

\bigskip
\noindent{\bf Acknowledgement:}

This research is supported by the Research Grant Council of Hong Kong
through Project 402309.  Supports from Natural Science Foundation of
China, National Basic Research Program of China (973),
Shuguang and Innovation Program of Shanghai Education
Committee are acknowledged.  The computation is performed at
the Supercomputer Center of Shanghai and ECNU.

\medskip


\vfill\eject

\begin{table}
\caption{$E_{release}$, the barriers for a hydrogen molecule to diffuse
through an aqueous valve, for various internal pressure $P_{int}$ and
external pressure $P_{ext}$.}

\centering

\vskip 1cm

\begin{tabular}{ccccc}

\hline

 System & $P_{ext}(GPa)$ & $P_{int}(GPa)$ & $E_{release}(eV) $ \\

\hline

       $(12,12)-COOH$& 0 &  0 & 0.85$\pm$0.23  \\
                     & 0 &  1 & 0.79$\pm$0.16  \\
                     & 0 &  2 & 0.81$\pm$0.16  \\
                     & 2 &  0 & 1.08$\pm$0.30  \\
                     & 2 &  2 & 0.95$\pm$0.07  \\

       \hline

       $(15,15)-COOH$& 0 &  0 & 0.67$\pm$0.17  \\
                     & 0 &  1 & 0.64$\pm$0.20  \\
                     & 0 &  2 & 0.32$\pm$0.08  \\
                     & 2 &  0 & 0.91$\pm$0.11  \\
                     & 2 &  2 & 0.80$\pm$0.17  \\

\end{tabular}

\end{table}

\vfill\eject

\begin{figure}
\centering
\includegraphics[width=5in]{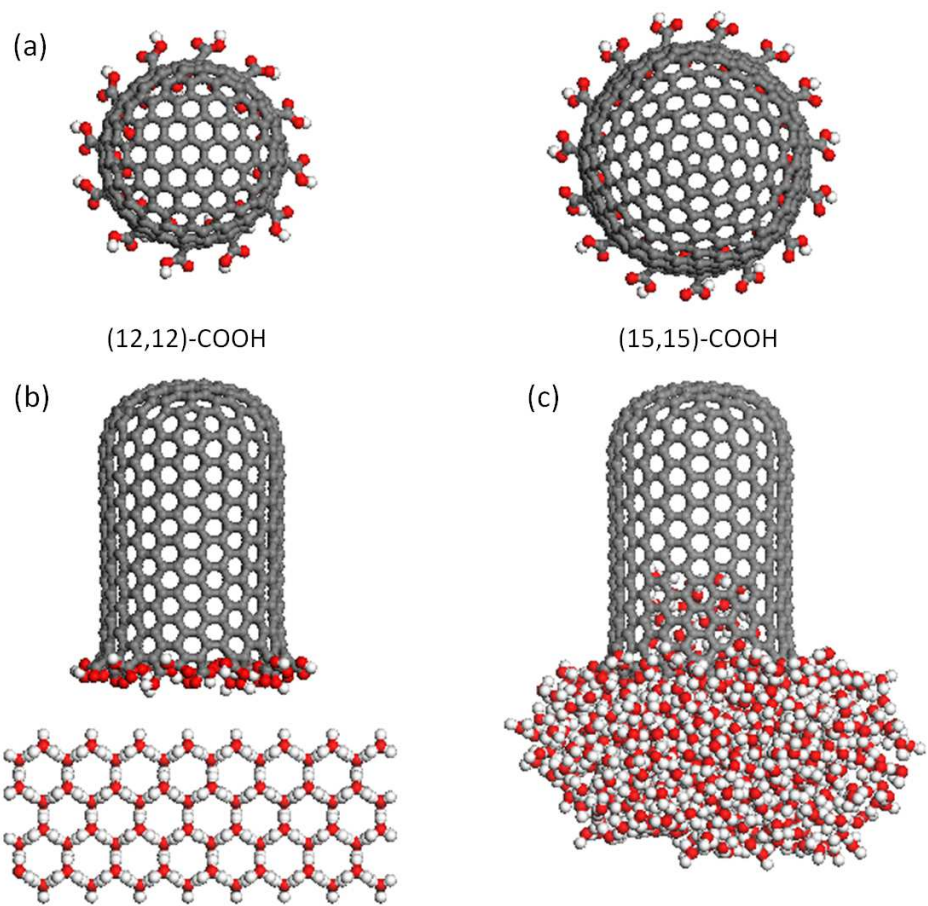}
\caption{\label{Scheme1} Formation of an aqueous valve
in the absence of hydrogen. (a) Top view of valve (12,12)-COOH and (15,15)-COOH;
(b)A nano-container separated from a chunk of ice, as the initial configuration
for MD simulation; (c)Melting of the ice chuck leads to the self-assembly
of water molecules around the end -COOH groups.}
\end{figure}

\vfill\eject

\begin{figure}
\centering
\includegraphics[width=5in]{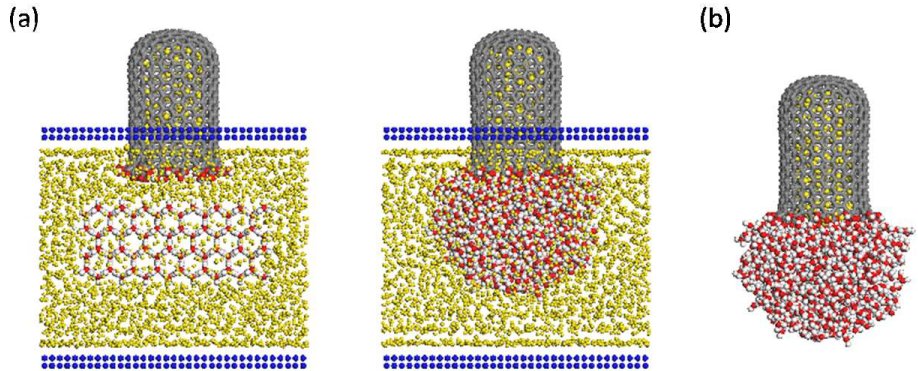}
\caption{\label{Scheme2} Formation of an aqueous valve
in the presence of hydrogen with a pressure of 0.6 GPa
which is controlled by two slabs (in blue).
(a) In the initial structure, the ice chunk is again frozen,
while hydrogen molecules (yellow particles) of 0.6 GPa fill
the container and and the ice lattice.  After melting
of the ice chunk, water molecules again self-assemble
around the -COOH groups to form an aqueous vavle.
(b) After the withdrawal of external hydrogen, the
aqueous valve is stable.  It demonstrates a filling
procedure in which the hydrogen is filled into open
container in high pressure and afterwards the container
is sealed off by exposure to liquid water.}
\end{figure}

\vfill\eject

\begin{figure}
\centering
\includegraphics[width=5in]{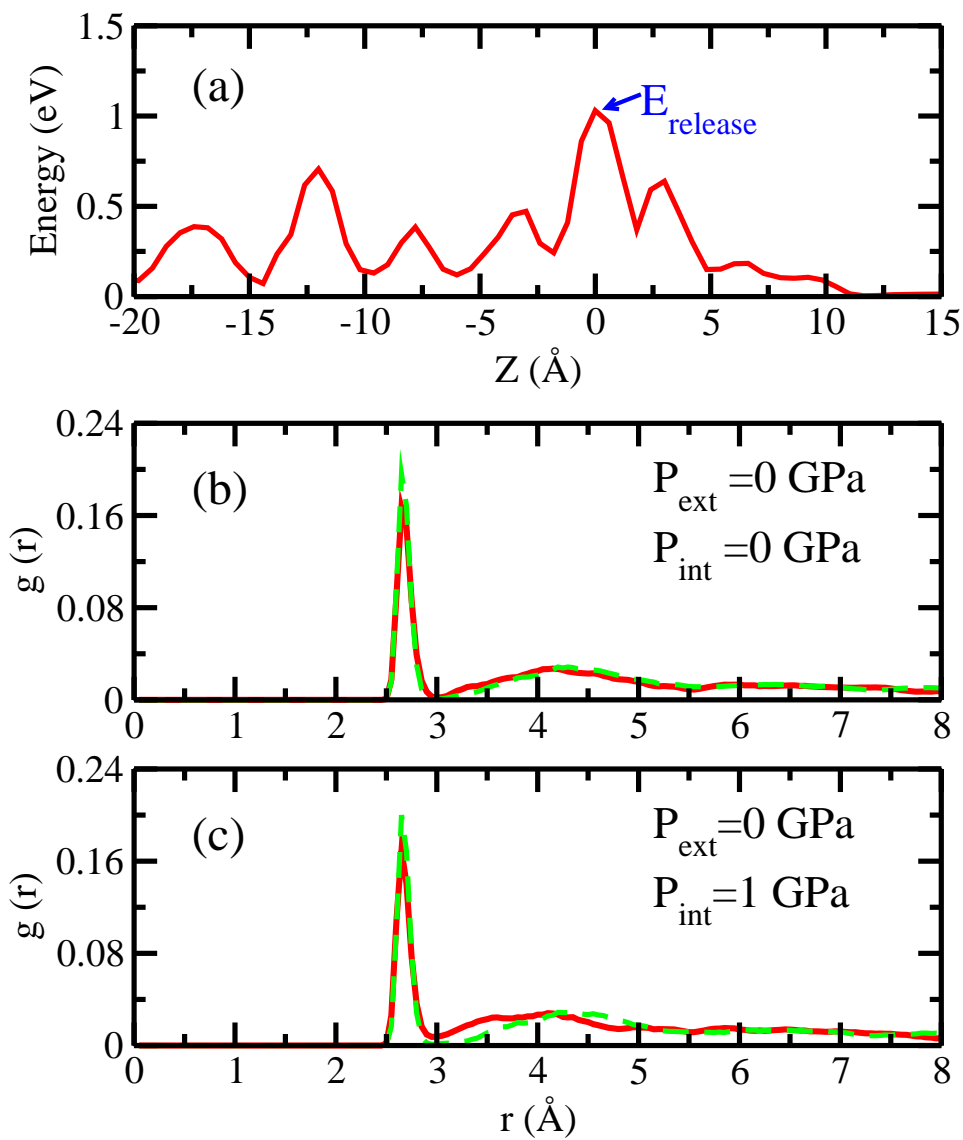}
\caption{\label{Scheme3} (a) A typical potential
energy curve as a hydrogen molecule diffuses
through the valve. (b) Radial distribution function {\it g(r)}
for O--O in water close to(red)/far from(green) the carboxylic
groups for P$_{ext}$=0 and P$_{int}$=0 GPa.  (c) {\it g(r)}
for O--O for P$_{ext}$=0 and P$_{int}$=1 GPa. }
\end{figure}

\vfill\eject

\begin{figure}
\centering
\includegraphics[keepaspectratio=true, scale = 0.80]{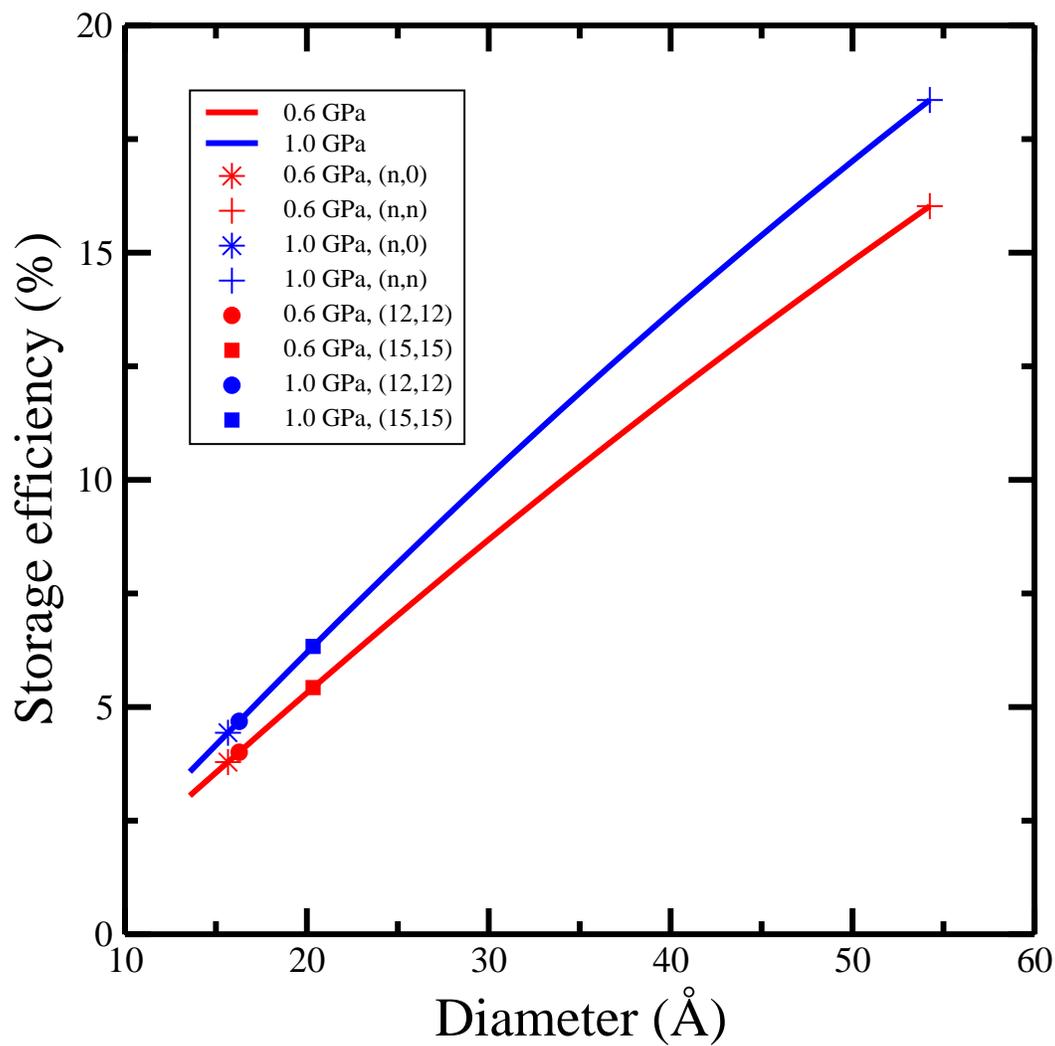}
\caption{\label{Scheme4} Calculated weight efficiency
curve as a function of tube diameter, under two conditions,
0.6 GPa/273 K, and 1.0 GPa/300K. }
\end{figure}

\end{document}